\RequirePackage{ifpdf}
\ifpdf % We are running pdfTeX in pdf mode
\documentclass[pdftex]{sigma}
\else
\documentclass{sigma}
\fi

\numberwithin{equation}{section}

\begin{document}

\allowdisplaybreaks

\renewcommand{\PaperNumber}{015}

\FirstPageHeading

\renewcommand{\thefootnote}{$\star$}

\ShortArticleName{KP Trigonometric Solitons and an Adelic Flag
Manifold}

\ArticleName{KP Trigonometric Solitons \\and an Adelic Flag
Manifold\footnote{This paper is a contribution to the Vadim
Kuznetsov Memorial Issue ``Integrable Systems and Related
Topics''. The full collection is available at
\href{http://www.emis.de/journals/SIGMA/kuznetsov.html}{http://www.emis.de/journals/SIGMA/kuznetsov.html}}}

\Author{Luc  HAINE} \AuthorNameForHeading{L. Haine}
\Address{Department of Mathematics, Universit\'e catholique de Louvain,\\
Chemin du Cyclotron 2, 1348 Louvain-la-Neuve, Belgium}
\Email{\href{mailto:haine@math.ucl.ac.be}{haine@math.ucl.ac.be}}

\ArticleDates{Received November 22, 2006, in f\/inal form January
5, 2007; Published online January 27, 2007}

\Abstract{We show that the trigonometric solitons of the KP
hierarchy enjoy a dif\/ferential-dif\/ference bispectral property,
which becomes transparent when translated on two suitable spaces
of pairs of matrices satisfying certain rank one conditions. The
result can be seen as a non-self-dual illustration of Wilson's
fundamental idea [{\it Invent. Math.} \textbf{133} (1998), 1--41]
for understanding the (self-dual) bispectral property of the
rational solutions of the KP hierarchy. It also gives a bispectral
interpretation of a (dynamical) duality between the hyperbolic
Calogero--Moser system and the rational Ruijsenaars--Schneider
system, which was f\/irst observed by Ruijsenaars~[{\it Comm.
Math. Phys.} \textbf{115} (1988), 127--165].}

\Keywords{Calogero--Moser type systems; bispectral problems}

\Classification{35Q53; 37K10}

\begin{flushright}
\emph{Dedicated to the memory of Vadim Kuznetsov}
\end{flushright}

\renewcommand{\thefootnote}{\arabic{footnote}}
\setcounter{footnote}{0}

\section{Introduction}
\looseness=1 One of the many gems I had the chance to share with
Vadim Kuznetsov during his visit to Louvain-la-Neuve in the fall
semester of 2000, had to do with his work on separation of
variables and spectrality. He knew about my work on bispectral
problems and insisted that these problems were connected. He
introduced me to his work with Nijhof\/f and Sklyanin~\cite{KNS},
on separation of variables for the elliptic Calogero--Moser
system. I would have loved to discuss the topic of the present
paper with him, which deals with the (simpler) trigonometric
version of this system.

One of the exciting new developments in the f\/ield of integrable
systems has been the introduction
 by George Wilson \cite{W2} of the so called Calogero--Moser spaces
\begin{gather}
\label{1.1} C_{N}=\big\{(X,Z)\in gl(N,\mathbb{C})\times
gl(N,\mathbb{C})\colon\mbox{rank}([X,Z]+I)=1\big\}/GL(N,\mathbb{C}),
\end{gather}
where $gl(N,\mathbb{C})$ denotes the space of complex $N\times N$
matrices, $I$ is the identity matrix and the complex linear group
$GL(N,\mathbb{C})$ acts by simultaneous conjugation of $X$ and
$Z$. These spaces are at the crossroads of many areas in
mathematics, connecting with such f\/ields as non-commutative
algebraic and symplectic geometry. For an introduction as well as
a broad overview of the subject, I recommend Etingof's recent
lectures \cite{E} at ETH (Z\"urich). Since the bispectral problem
is not mentioned in these lectures, it seems not inappropriate to
start from this problem as introduced in the seminal paper
\cite{DG} by Duistermaat and Gr\"unbaum, which has played and (as
we shall see) continues to play a decisive role in the subject.

In the form discussed by Wilson \cite{W1}, the bispectral problem
asks for the classif\/ication of all rank 1 commutative algebras
$\mathcal{A}$ of dif\/ferential operators, for which the joint
eigenfunction $\psi(x,z)$ which satisf\/ies
\begin{gather}\label{1.2}
A(x,\partial/\partial x)\psi(x,z)=f_{A}(z)\psi(x,z)\qquad \forall
\,A\in\mathcal{A},
\end{gather}
also satisf\/ies a (non-trivial) dif\/ferential equation in the
spectral variable
\begin{gather}\label{1.3}
B(z,\partial/\partial z)\psi(x,z)=g(x)\psi(x,z).
\end{gather}
In \cite{W1}, it was found that all the solutions of the problem
are parametrized by a certain subgrassmannian of the Segal--Wilson
Grassmannian ${\rm Gr}$, that Wilson called the adelic Grassmannian and
that he denoted by ${\rm Gr}^{\rm ad}$. The same Grassmannian
parametrizes the rational solutions in $x$ of the
Kadomtsev--Petviashvili (KP) equation  (vanishing as
$x\to\infty$). The main result of~\cite{W2} is to give another
description of ${\rm Gr}^{\rm ad}$ as the union $\cup_{N\geq 0}
C_{N}$ of the Calogero--Moser spaces introduced above. The
correspondence can be seen as given by the map
\begin{gather}\label{1.4}
\beta\colon (X,Z) \to \psi_W(x,z)=e^{xz}\;\det
\{I-(xI-X)^{-1}(zI-Z)^{-1}\},
\end{gather}
which sends a pair $(X,Z)\in C_N$ (modulo conjugation) to the
(stationary) Baker--Akhiezer function of the corresponding space
$W=\beta(X,Z)\in {\rm Gr}^{\rm ad}$. From \eqref{1.4}, one sees
immediately that the mysterious bispectral involution $b\colon
{\rm Gr}^{\rm ad}\to {\rm Gr}^{\rm ad}$ which exchanges the role
of the variables~$x$ and~$z$
\begin{gather*}
\psi_{b(W)}(x,z)=\psi_W(z,x),\qquad W\in {\rm Gr}^{\rm ad},
\end{gather*}
becomes transparent when expressed at the level of the
Calogero--Moser spaces, as it is given by $b^{C}(X,Z)=(Z^t,X^t)$,
where $X^t$ and $Z^t$ are the transposes of $X$ and $Z$. The
situation is nicely summarized by the following commutative
diagram, all arrows of which are bijections
\begin{gather}\label{1.5}
\begin{CD}
\cup_{N\geq 0}C_N @>\beta >>{\rm Gr}^{\rm ad}\\
@VV{b^{C}}V             @VV{b}V\\
\cup_{N\geq 0}C_N @>\beta>>{\rm Gr}^{\rm ad}
\end{CD}
\end{gather}

In \cite{HI}, jointly with Plamen Iliev, we considered the
following discrete-continuous version of the bispectral problem.
To determine all rank 1 commutative algebras $\mathcal{A}$ of
dif\/ference operators, for which the joint eigenfunction
$\psi(n,z)$ which satisf\/ies
\begin{gather}\label{1.6}
A\psi(n,z)\equiv \sum_{{\rm finitely \
many}\;j\in\mathbb{Z}}a_j(n)\psi(n+j,z)=f_A(z)\psi(n,z)\qquad
\forall \, A\in\mathcal{A},
\end{gather}
also satisf\/ies a (non-trivial) dif\/ferential equation in the
spectral variable
\begin{gather}\label{1.7}
B(z,\partial/\partial z)\psi(n,z)=g(n)\psi(n,z).
\end{gather}
The problem was motivated by an earlier work with Alberto
Gr\"unbaum \cite{GH}, where we investigated the situation when the
algebra $\mathcal{A}$ contains a second-order symmetric
dif\/ference operator (this time, without imposing any rank
condition on $\mathcal{A}$). This situation extends the theory of
the classical orthogonal polynomials, where the dif\/ferential
equation is of the second order too.

The main result of \cite{HI} was to construct from Wilson's adelic
Grassmannian ${\rm Gr}^{\rm ad}$ an (isomorphic) adelic f\/lag
manifold ${\rm Fl}^{\rm ad}$, which provides solutions of the
bispectral problem raised above, and parametrizes rational
solutions in $n$ (vanishing as $n\to\infty$) of the discrete KP
hierarchy. The message of this paper is to show that the analogue
of Wilson's diagram \eqref{1.5} in the context of this new
bispectral problem is the following commutative diagram, with
bijective arrows
\begin{gather}\label{1.8}
\begin{CD}
\cup_{N\geq 0}C_N @>\beta >>{\rm Gr}^{\rm ad}\equiv {\rm Fl}^{\rm ad}\\
@VV{b^C}V                    @VV{b}V\\
\cup_{N\geq 0}C_N^{\rm trig}@>\beta^{\rm trig}>>{\rm Gr}^{\rm
trig}
\end{CD}
\end{gather}
In this diagram, the spaces
\begin{gather}\label{1.9}
C_{N}^{\rm trig}=\big\{(X,Z)\in GL(N,\mathbb{C})\times
gl(N,\mathbb{C})\colon \mbox{rank}\;
(XZX^{-1}-Z+I)=1\big\}/GL(N,\mathbb{C}),
\end{gather}
are trigonometric analogues of the Calogero--Moser spaces $C_N$
def\/ined in \eqref{1.1}. ${\rm Gr}^{\rm trig}$ is a certain
subgrassmannian of linear spaces $W\in {\rm Gr}$ parametrizing
special solitons of the KP hierarchy, that I call the
``trigonometric Grassmannian'', since the corresponding tau
functions $\tau_W$ take the form
\begin{gather}\label{1.10}
\tau_W(x+t_1,t_2,t_3,\ldots)=\prod_{i=1}^{N}2\sinh
\frac{\big(x-x_i(t_1,t_2,t_3,\ldots)\big)}{2},
\end{gather}
with $x_i(t_1,t_2,t_3,\ldots)$ being a solution of the
trigonometric Calogero--Moser--Sutherland hierarchy (as long as
all $x_i(t_1,t_2,t_3,\ldots)$ remain distinct, see Section 3). The
bispectral map $b\colon {\rm Fl}^{\rm ad}\to {\rm Gr}^{\rm trig}$,
which sends a f\/lag $\mathcal{V}$ to a linear space
$W=b(\mathcal{V})$, and is def\/ined by
\begin{gather}\label{1.11}
\psi_{b(\mathcal{V})}(x,z)=\psi_{\mathcal{V}}(z,e^x-1),
\end{gather}
trivializes when expressed at the level of the Calogero--Moser
spaces, as it is now given by\footnote[1]{The additional condition
$\det (I+Z)\neq 0$ needed for this def\/inition to make sense,
follows from f\/ixing the radius of the circle used in the
def\/inition of ${\rm Gr}$ to be $1$, which can always be assumed,
see Section 4.}
\begin{gather}\label{1.12}
b^{C}(X,Z)=\big(I+Z^{t},X^{t}(I+Z^t)\big).
\end{gather}
The result will follow easily from the expression of the
(stationary) Baker--Akhiezer function $\psi_W(x,z)$ of a space
$W\in {\rm Gr}^{\rm trig}$, in terms of pairs of matrices
$(X,Z)\in C_N^{\rm trig}$
\begin{gather}\label{1.13}
\psi_W(x,z)=e^{xz} \det \{I-X(e^{x}I-X)^{-1}(zI-Z)^{-1}\},
\end{gather}
which def\/ines the map $\beta^{\rm trig}$ in the diagram
\eqref{1.8}. The def\/inition of the map $\beta$ in the same
diagram follows immediately from the def\/inition of the adelic
f\/lag manifold in terms of Wilson's adelic Grassmannian as given
in \cite{HI}, and will be recalled in Section 4.

Some time ago, Ruijsenaars \cite{R1} (see also \cite{R2}) made a
thorough study of the action-angle maps for Calogero--Moser type
systems with repulsive potentials, via the study of their
scattering theory. Along the way, he observed various duality
relations between these systems. In particular, when the
interaction between the particles in the trigonometric
Calogero--Moser system is repulsive, the system is dual (in the
sense of scattering theory) to the rational Ruijsenaars--Schneider
system. In the last section, we show that within our picture
\eqref{1.8}, if $\tau_W(t_1,t_2,t_3,\ldots)$ is the tau function
of a space $W\in {\rm Gr}^{\rm trig}$ as in \eqref{1.10}, the tau
function of the f\/lag $b^{-1}(W)$ is
\begin{gather*}
\tau
_{b^{-1}(W)}(n,t_1,t_2,\ldots)=\prod_{i=1}^{N}(n-\lambda_i(t_1,t_2,\ldots)),\qquad
n\in\mathbb{Z},
\end{gather*}
with $\lambda_i(t_1,t_2,\ldots)$ solving now the rational
Ruijsenaars--Schneider hierarchy, thus representing Ruijsenaars'
duality as a bispectral map. The likelihood of this last statement
was formulated previously by Kasman \cite{K}, on the basis of a
similar relationship between the quantum versions of these
systems, as studied by Chalykh \cite{C1} (see also \cite{C2}). It
is also implicitly suggested by a recent work of Iliev \cite{I},
which relates the polynomial tau functions in $n$ of the discrete
KP hierarchy with the rational Ruijsenaars--Schneider hierarchy.
%%%%%%%%%%%%%%%%%%%%%%%%%%%%%%%%%%%%%%%%%%%%%%%%%%%%%%%%%%%%%%%%%%%%%%%%%%%%%%%%%%%%%%%%%%%%%%%%%%%%%%%%%%%%% Section 2
\section{Trigonometric solitons of the KP hierarchy}
In this section, we construct a class of special solitons of the
KP hierarchy. Their relation with the trigonometric version of the
Calogero--Moser hierarchy will be explained in the next section,
justifying the appellation ``trigonometric solitons''. We f\/irst
need to recall brief\/ly the def\/inition of the Segal--Wilson
Grassmannian ${\rm Gr}$ and its subgrassmannian ${\rm Gr}^{\rm
rat}$, from which solitonic solutions of the KP hierarchy can be
constructed, see \cite{SW} for details. Let $S^1\subset\mathbb{C}$
be the unit circle, with center the origin, and let $H$ denote the
Hilbert space $L^2(S^1,\mathbb{C})$. We split~$H$ as the
orthogonal direct sum $H=H_{+}\oplus H_{-}$, where $H_{+}$
(resp.~$H_{-}$) consists of the functions whose Fourier series
involves only non-negative (resp. only negative) powers of $z$.
Then ${\rm Gr}$ is the Grassmannian of all closed subspaces $W$ of
$H$ such that (i) the projection $W\to H_{+}$ is a Fredholm
operator of index zero (hence generically an isomorphism); (ii)
the projection $W\to H_{-}$ is a compact operator. For
$t=(t_1,t_2,\ldots)$, let
$\exp(t,z)=\exp(\sum\limits_{k=1}^{\infty}t_k z^k)$. For all $t$,
$\exp^{-1}(t,z)W$ belongs to ${\rm Gr}$ and, for almost any $t$, it is
isomorphic to $H_+$, so that there is a unique function in it
$\tilde{\psi}_W(t,z)$ which projects onto $1$. The function
$\psi_W(t,z)=\exp(t,z)\tilde{\psi}_W(t,z)$ is called the
Baker--Akhiezer function of the space $W$ and
$\tilde{\psi}_W(t,z)$ is called the reduced Baker--Akhiezer
function. A fundamental result of Sato asserts that there is a
unique (up to multiplication by a~constant) function
$\tau_W(t_1,t_2,t_3,\ldots)$, the celebrated tau function, such
that
\begin{gather}\label{2.1}
\psi_W(t,z)=\exp(t,z)\frac{\tau_W(t_1-{1}/{z},t_2-{1}/(2z^2),t_3-{1}/(3z^3),\ldots)}
{\tau_W(t_1,t_2,t_3,\ldots)}.
\end{gather}
An element of the form $\sum\limits_{k\leq s}a_k z^k$, $a_s\neq
0$, is called an element of f\/inite order $s$. For $W\in {\rm
Gr}$,  $W^{\rm alg}$ denotes the subspace of elements of f\/inite
order of $W$. It is a dense subspace of $W$. We also need the ring
\begin{gather}\label{2.2}
A_W=\{f\;\mbox{analytic in a neighborhood of}\;S^1\colon f.W^{\rm
alg}\subset W^{\rm alg}\}.
\end{gather}
The rational Grassmannian ${\rm Gr}^{\rm rat}$ is the subset of
$\rm Gr$, for which $\mathop{\rm Spec}(A_W)$ is a rational curve.
In this case, $A_W$ is a subset of the ring $\mathbb{C}[z]$ of
polynomials in $z$ and the map $\mathbb{C}\to\mathop{\rm
Spec}(A_W)$ induced by the inclusion is a birational isomorphism,
sending $\infty$ to a smooth point completing the curve.

Let us f\/ix $N$ distinct complex numbers
$\lambda_1,\ldots,\lambda_N$ inside of $S^1$, and another $N$
non-zero complex numbers $\mu_1,\ldots,\mu_N$. We assume that
$\lambda_i-\lambda_j\neq 1,\forall\;i\neq j$. We def\/ine $W^{\rm
alg}_{\lambda,\mu}$ as the space of rational functions $f(z)$ such
that

(i) $f$ is regular except for (at most) simple  poles at
$\lambda_1,\ldots,\lambda_N$ and poles of any order at inf\/inity;

(ii) $f$ satisf\/ies the $N$ conditions
\begin{gather}\label{2.3}
{\rm res}_{\lambda_i}f(z)+\mu_i^2 f(\lambda_i-1)=0,\qquad 1\leq
i\leq N,
\end{gather}
where ${\rm res}_{\lambda_i}f(z)$ is the residue of $f(z)$ at
$\lambda_i$. The closure of $W^{\rm alg}_{\lambda,\mu}$ in
$L^2(S^1,\mathbb{C})$ def\/ines a space $W_{\lambda,\mu}\in {\rm
Gr}^{\rm rat}$.

The Baker--Akhiezer function $\psi_{W}$ of a space
$W=W_{\lambda,\mu}$ has the form
\begin{gather}\label{2.4}
\psi_{W}(t,z)=\exp(t,z)\Biggl\{1-\sum_{j=1}^{N}\frac{\mu_j
b_j(t)}{z-\lambda_j}\Biggr\},
\end{gather}
for some functions $b_j(t)$ determined by \eqref{2.3}. By a simple
computation, we obtain
\begin{gather}\label{2.5}
-\exp(t,\lambda_i)b_i(t)+\mu_i\exp(t,\lambda_i-1)\Biggl\{1+\sum_{j=1}^{N}\frac{\mu_jb_j(t)}
{1+\lambda_j-\lambda_i}\Biggr\}=0.
\end{gather}
We introduce the $N\times N$ matrices $X$ and $Z$ with entries
\begin{gather}\label{2.6}
X_{ij}=\frac{\mu_i\mu_j}{1+\lambda_j-\lambda_i},\qquad
Z_{ij}=\lambda_i\delta_{ij},
\end{gather}
with $\delta_{ij}$ the usual Kronecker symbol so that $Z$ is
diagonal, and we put
\begin{gather}\label{2.7}
\tilde{X}=\exp\Biggl\{\sum_{k=1}^{\infty}t_k(Z^k-(Z-I)^k)\Biggr\}-X.
\end{gather}
With these notations, the system of equations \eqref{2.5}
determining the functions $b_i(t)$ is written as
\begin{gather}\label{2.8}
\tilde{X}b(t)=\mu,
\end{gather}
where $b(t)$ and $\mu$ denote column vectors of length $N$, with
entries $b_i(t)$ and $\mu_i$ respectively. From \eqref{2.4} and
\eqref{2.8}, we get that the reduced Baker--Akhiezer function
$\tilde{\psi}_W(t,z)=\exp^{-1}(t,z)\psi_W(t,z)$ can be written as
\begin{gather}\label{2.9}
\tilde{\psi}_W(t,z)=1-\mu^t\big(zI-Z\big)^{-1}\tilde{X}^{-1}\mu,
\end{gather}
with $\mu^t$ the row vector obtained by transposing $\mu$.

The following commutation relation will be crucial
\begin{gather}\label{2.10}
[X,Z]=\mu\mu^t-X.
\end{gather}
For short set
\begin{gather}\label{2.11}
\tilde{Z} =zI-Z.
\end{gather}
Considering that if $T$ is a matrix of rank $1$ then
$1+\mathop{\rm tr}T=\det (I+T)$ (with tr denoting the trace), from
\eqref{2.7}, \eqref{2.9}, \eqref{2.10} and \eqref{2.11}, we f\/ind
\begin{gather*}
\tilde{\psi}_W(t,z)=1-\mathop{\rm
tr}\big\{\tilde{X}^{-1}\mu\mu^t\tilde{Z}^{-1}\big\} =\det
\big\{I-\tilde{X}^{-1}([X,Z]+X)\tilde{Z}^{-1}\big\}
\\
\phantom{\tilde{\psi}_W(t,z)}{} {}=\det
\big\{I-\tilde{X}^{-1}([\tilde{X},\tilde{Z}]+X)\tilde{Z}^{-1}\big\}
=\det
\big\{\tilde{X}^{-1}(\tilde{Z}\tilde{X}-X)\tilde{Z}^{-1}\big\}.
\end{gather*}
Using the fact that the determinant of a product of matrices does
not depend on the orders of the factors, we get
\begin{gather}\label{2.12}
\tilde{\psi}_W(t,z)=\det
\big\{I-X\tilde{X}^{-1}\tilde{Z}^{-1}\big\}.
\end{gather}
In particular, setting $t_2=t_3=\cdots=0$ and $t_1=x$, we f\/ind
that the so-called stationary Baker--Akhiezer function of $W$
admits the following form
\begin{gather}\label{2.13}
\psi_W(x,z)=e^{xz} \det \big\{I-X(e^{x}I-X)^{-1}(zI-Z)^{-1}\big\}.
\end{gather}
By Cauchy's determinant formula applied to $X$ in \eqref{2.6}
\begin{gather}\label{2.14}
\det (X)=\prod_{i=1}^{N}\mu_i^2\prod_{1\leq i<j\leq
N}\biggl(1-\frac{1}{1-(\lambda_i-\lambda_j)^2}\biggr),
\end{gather}
which is non-zero by the assumptions we made on the $\lambda_j$'s
and $\mu_j$'s. Thus $X$ is invertible and it follows from
\eqref{2.10} that $\mbox{rank}(XZX^{-1}-Z+I)=1$, as announced in
the Introduction, see~\eqref{1.9} and \eqref{1.13}. It is now easy
to deduce the following proposition.
\begin{proposition}\label{proposition 1}
The tau function $\tau_W(t_1,t_2,t_3,\ldots)$ of a space
$W=W_{\lambda,\mu}$ is given by
\begin{gather}\label{2.15}
\tau_W(t_1,t_2,t_3,\ldots)=\det
\Biggl\{I-X\exp\Biggl\{\sum_{k=1}^{\infty}t_k\big((Z-I)^k-Z^k\big)\Biggr\}\Biggr\},
\end{gather}
with $X$ and $Z$ defined as in \eqref{2.6}.
\end{proposition}
\begin{proof} Denoting for short by $\exp\{\cdots\}$ the expression that
appears inside the exponential in~\eqref{2.15}, one computes
\begin{gather*}
\tau_W\biggl(t_1-\frac{1}{z},t_2-\frac{1}{2z^2},t_3-\frac{1}{3z^3},\ldots\biggr)
=\det \{I-X\exp\{\cdots\}(zI-(Z-I))(zI-Z)^{-1}\}
\\ \qquad
=\det \{I-X\exp\{\cdots\}-X\exp\{\cdots\}(zI-Z)^{-1}\},
\end{gather*}
from which it follows that
\begin{gather*}
\frac{\tau_W(t_1-{1}/{z},t_2-{1}/({2z^2}),t_3-{1}/({3z^3}),\ldots)}
{\tau_W(t_1,t_2,t_3,\ldots)}=\det
\{I-X(\exp^{-1}\{\ldots\}-X)^{-1}(zI-Z)^{-1}\}
\\ \qquad
=\det \{I-X\tilde{X}^{-1}\tilde{Z}^{-1}\},
\end{gather*}
with $\tilde{X}$ and $\tilde{Z}$ def\/ined as in \eqref{2.7} and
\eqref{2.11}. This shows that the reduced Baker--Akhiezer function
obtained in \eqref{2.12} satisf\/ies Sato's formula \eqref{2.1}
with $\tau_W$ as in \eqref{2.15}. Since this formula determines
the tau function up to a constant, the proof is complete.
\end{proof}

\begin{remark}\hspace{-1mm}\footnote[2]{I am thankful to the two referees
both of whom made this important observation.}\label{remark 1}
Kasman and Gekhtman \cite{KG} (see Corollary 3.2 in their paper)
have established that for any triple $(X,Y,Z)$ of $N\times N$
matrices such that rank $(XZ-YX)=1$, the function
\begin{gather}\label{2.16}
\tau_{(X,Y,Z)}(t_1,t_2,\ldots)=\det
\Biggl\{I-X\exp\Biggl\{-\sum_{k=1}^{\infty}t_k
Z^k\Biggr\}\exp\Biggl\{\sum_{k=1}^{\infty}t_k Y^k\Biggr\}\Biggr\},
\end{gather}
is a tau function of the KP hierarchy, associated with some $W\in
{\rm Gr}^{\rm rat}$, by showing that it satisf\/ies the Hirota
equation in Miwa form. The special choice
\begin{gather*}
X_{ij}=\frac{\mu_i\mu_j}{\lambda_j-\nu_i},\qquad
Y_{ij}=\nu_i\delta_{ij},\qquad Z_{ij}=\lambda_i\delta_{ij}, \qquad
\mbox{with}\quad \nu_i\neq\lambda_j,\qquad \forall \;i,j,
\end{gather*}
leads to a $N$-soliton solution of the KP hierarchy, and
Proposition~\ref{proposition 1} can be obtained by picking
$Y=Z-I$.
\end{remark}

Thanks to Remark~\ref{remark 1}, it makes sense to introduce the
following def\/inition:
\begin{definition}\label{definition 1}
The trigonometric Grassmannian ${\rm Gr}^{\rm trig}$ is def\/ined
to be the following subgrassmannian of  ${\rm Gr}^{\rm rat}$
\begin{gather}\label{2.17}
{\rm Gr}^{\rm trig}\!=\!\big\{W\!\in\! {\rm Gr}^{\rm rat}\colon
\tau_W(t_1,t_2,\ldots)\!
=\!\tau_{(X,Z-I,Z)}(t_1,t_2,\ldots),\mbox{for}\;(X,Z)\! \in\!
\textstyle\bigcup_{N\geq 0}C_N^{\rm trig}\big\},
\end{gather}
with $C_N^{\rm trig}$ def\/ined as in \eqref{1.9} and with
$\tau_{(X,Z-I,Z)}(t_1,t_2,\ldots)$ def\/ined as in \eqref{2.16}.
The corresponding solutions of the KP hierarchy will be called
\emph{trigonometric solitons}.
\end{definition}
\begin{example} As an example of a non-generic $W\in {\rm Gr}^{\rm trig}$
(i.e. not of the form $W_{\lambda,\mu}$ as above), let us def\/ine
$W\in {\rm Gr}^{\rm rat}$ to be the closure in
$L^2(S^1,\mathbb{C})$ of the space $W^{\rm alg}$ formed with the
rational functions $f(z)$ such that

(i) $f$ is regular except for (at most) a double pole at $z=0$ and
a pole of any order at $\infty$;

(ii) $f$ satisf\/ies the two conditions
\begin{gather*}
{\rm res}_0 zf(z)+f(-1)=0,
\\
{\rm res}_0 f(z)+f'(-1)=0.
\end{gather*}
A simple computation shows that the corresponding stationary
Baker--Akhiezer function is
\begin{gather*}
\psi_W(x,z)=e^{xz}\biggl\{1+\frac{2}{(1+e^{2x})z}+\frac{1-e^x}{(1+e^{2x})z^2}\biggr\},
\end{gather*}
which can be put into the form \eqref{2.13} with
\begin{gather*}
X=\begin{pmatrix} 0&-1
\\1&0\end{pmatrix}, \qquad
Z=\begin{pmatrix}0&1
\\0&0
\end{pmatrix},
\end{gather*}
which forms a trigonometric Calogero--Moser pair as def\/ined in
\eqref{1.9}, with a non-diagonali\-zab\-le~$Z$. The corresponding
tau function is
\begin{gather*}
\tau_W(x,t_2,t_3,\ldots)\!=\!e^{-2x}\Biggl\{\!\exp\Biggl(\!2\sum_{k=2}^{\infty}(-1)^kt_k\!\Biggr)
\!-\!\!\exp\Biggl(\sum_{k=2}^{\infty}(-1)^kt_k\!\Biggr)\!
\Biggl(\sum_{k=2}^{\infty}(-1)^{k+1}kt_k\!\Biggr)e^{x}\!+\!e^{2x}\!\Biggr\}.
\end{gather*}
\end{example}

\section[${{\rm Gr}}^{\rm \mathbf{trig}}$ and the Calogero-Moser-Sutherland
hierarchy]{$\boldsymbol{{{\rm Gr}}^{\rm \mathbf{trig}}}$ and the
Calogero--Moser--Sutherland hierarchy}

In this section, we relate the Grassmannian ${\rm Gr}^{\rm trig}$
introduced in Def\/inition~\ref{definition 1} with the
Calogero--Moser--Sutherland system, also referred as the
trigonometric (or hyperbolic) Calogero--Moser system. This is a
system of $N$ particles on the line whose motion is governed by
the Hamiltonian
\begin{gather}\label{3.1}
H(x,y)=\frac{1}{2}\sum_{i=1}^{N}y_i^2-\sum_{1\leq i<j\leq N}
\frac{1}{4\sinh^2((1/2)(x_i-x_j))}.
\end{gather}
We allow the particles to move in the complex plane. The original
Moser--Sutherland system \cite{M} is recovered if we suppose our
particles to be conf\/ined to the imaginary axis, since the
hyperbolic sine becomes then the trigonometric one. When the
motion of the particles is conf\/ined to the real axis, the
potential is attractive if velocities are real, and repulsive if
velocities (and time) are purely imaginary. In this last case, the
particles ultimately behave like free particles. But for most
initial conditions in the complex plane, some collisions will take
place after a f\/inite time.

{\sloppy Moser \cite{M} proved that the system \eqref{3.1} is
completely integrable, showing that it describes an isospectral
deformation of  the $N\times N$ matrix $L(x,y)$ with entries
\begin{gather}\label{3.2}
L_{ij}(x,y)=\delta_{ij}y_i+(1-\delta_{ij})\frac{1}{2\sinh((1/2)(x_i-x_j))},
\end{gather}
where $\delta_{ij}$ is the usual Kronecker symbol. More precisely,
the quantities $F_{k}(x,y)=(1/k)\mathop{\rm tr}L^k(x,y)$ (with tr
denoting the trace), $k=1,2,\ldots,N$, are $N$ independent f\/irst
integrals in involution for the system. In particular, $F_2(x,y)$
gives back the original Hamiltonian~$H(x,y)$.

}

In order to relate the system to the KP trigonometric solitons
introduced in Def\/inition~\ref{definition 1}, we need the
following lemma which can be extracted from Ruijsenaars \cite{R1},
and was motivated by his study of the scattering theory of the
system \eqref{3.1} when the interaction between the particles is
repulsive (which, with our conventions, amounts to pick velocities
and time imaginary).
\begin{lemma}\label{lemma 1}
Let $(X,Z)\in C_N^{\rm trig},N\geq 1$, as defined in \eqref{1.9}.
If $X$ is diagonalizable, there is a~conjugation
\begin{gather} \label{3.3}
U^{-1}XU=K^2(x),\qquad U^{-1}ZU=L(x,y),
\end{gather}
with $K(x)$ a diagonal matrix of the form
\begin{gather}\label{3.4}
K(x)={\rm diag}\big(e^{{x_1}/{2}},\ldots, e^{{x_N}/{2}}\big),
\end{gather}
and $L(x,y)$ as in \eqref{3.2}.
\end{lemma}
\begin{proof} Since $\det (X)\neq 0$, when diagonalizing $X$, we can always assume $K$
to have the form \eqref{3.4}. Denoting by $L$ the result of the
conjugation of $Z$ by the same matrix $U$, since by the
def\/inition of $C_N^{\rm trig}$ the rank of $[X,Z]+X$ is $1$, we
have
\begin{gather}\label{3.5}
[K^2,L]=\alpha\beta^t-K^2,
\end{gather}
with $\alpha$ and $\beta$ two (non-zero) column vectors of length
$N$. Writing \eqref{3.5} componentwise, we get
\begin{gather}\label{3.6}
(e^{x_i}-e^{x_j})L_{ij}=\alpha_i\beta_j-\delta_{ij}e^{x_i},
\end{gather}
which, by putting $i=j$, shows that $\alpha_i\beta_i=e^{x_i}\neq
0$, $\forall\, i$. Thus, by multiplying $U$ to the right by an
appropriate diagonal matrix, we can always arrange that
$\alpha_i=\beta_i=e^{x_i/2}$. With this choice, one sees from
\eqref{3.6} that necessarily $e^{x_i}\neq e^{x_j}$, $\forall \,
i\neq j$, and
\begin{gather*}
L_{ij}=\frac{1}{2\sinh\big((1/2)(x_i-x_j)\big)},\qquad
\mbox{for}\quad i\neq j,
\end{gather*}
while the diagonal entries $L_{ii}$ are free. Denoting
$L_{ii}=y_i$ establishes the lemma.
\end{proof}

The explicit integration of the system \eqref{3.1} was performed
by Olshanetsky and Perelo\-mov~\cite{OP}. An interpretation in
terms of Hamiltonian reduction was given by Kazhdan, Kostant and
Sternberg \cite{KKS}, leading to the following result whose proof
can be found in the nice treatise \cite{Su}, by Suris. We identify
$gl(N,\mathbb{C})$ with its dual via the trace form $\langle
X,Y\rangle =\mathop{\rm tr}(XY)$, and we def\/ine accordingly the
gradient of a smooth function $\varphi\colon
gl(N,\mathbb{C})\to\mathbb{C}$ by $\mbox{d}\varphi(X)(Y)=\langle
\nabla\varphi(X),Y\rangle$.
\begin{proposition}\label{proposition 2} \emph{(See \cite[Theorem 27.6]{Su} for a proof.)}
Let $\varphi\colon  gl(N,\mathbb{C})\to\mathbb{C}$ be an $\rm
Ad$-invariant function, and let $H(x,y)=\varphi(L(x,y))$, with
$L(x,y)$ as in \eqref{3.2}. Let $(x_i(t), y_i(t))$ be the solution
of Hamilton's equations
\begin{gather*}
\dot{x}_i=\frac{\partial H}{\partial y_i},\qquad
\dot{y}_i=-\frac{\partial H}{\partial x_i},
\end{gather*}
with initial conditions $(x_0,y_0)=\big(x_1(0),\ldots, x_N(0),
y_1(0),\ldots, y_N(0)\big)\in \mathbb{C}^{2N}$, such that
$x_i(0)\neq x_j(0)$, $\forall \, i\neq j$. Then, the quantities
$e^{x_i(t)}$ are the eigenvalues of the matrix
\begin{gather}\label{3.7}
K_0^2\exp\big(t\nabla \varphi(L_0)\big),
\end{gather}
with $K_0=K(x_0)$ as in \eqref{3.4}, $L_0=L(x_0,y_0)$ and $\nabla
\varphi$ the gradient of $\varphi$. Moreover, the mat\-rix~$V(t)$
which diagonalizes $K_0^2\exp\big(t\nabla \varphi(L_0)\big)$, so
that
\begin{gather*}
K^2(x(t))=V(t)K_0^2\exp\big(t\nabla \varphi(L_0)\big)V(t)^{-1},
\end{gather*}
and which is normalized by the condition
\begin{gather*}
V(t)K_0e=K(x(t))e,\qquad e=(1,\ldots,1)^t,
\end{gather*}
is such that
\begin{gather*}
L(x(t),y(t))=V(t)L_0 V(t)^{-1}.
\end{gather*}
\end{proposition}
\begin{remark} As explained at the beginning of this section, to describe
a repulsive interaction in the trigonometric Calogero--Moser
system with Hamiltonian $H(x,y)=\varphi\big(L(x,y)\big)$,
$\varphi(L)= (1/2)\mathop{\rm tr}L^2$ as in \eqref{3.1} and
\eqref{3.2}, we have to pick the $x_i$'s real, the $y_i$'s
imaginary and $t$ imaginary also. In this case,
$t\nabla\varphi(L_0)=tL_0,t\in\sqrt{-1}\;\mathbb{R}$, is hermitian
and thus $K_0^2\exp(tL_0)$ in \eqref{3.7} is always
diagonalizable. In the general case, this matrix can become
non-diagonalizable for some values of $t$, which leads to
collisions in the system.
\end{remark}
Let us now consider the following Hamiltonians $H_k(x,y)=\varphi_k
(L(x,y))$, with
\begin{gather}\label{3.8}
\varphi_k(L(x,y))=\frac{1}{k+1}\mathop{\rm
tr}\{(L(x,y)-I)^{k+1}-L^{k+1}(x,y)\},\qquad k=1,2,\ldots,
\end{gather}
and let us denote by $x_i(t)\equiv x_i(t_1,t_2,t_3,\ldots)$ the
solution obtained by f\/lowing along the Hamiltonian vector
f\/ield $X_{H_1}$ during a time $t_1$, $X_{H_2}$ during a time
$t_2$ etc., starting from some initial condition $(x_0,y_0)\in
\mathbb{C}^{2N}$. For short, we shall refer to $x_i(t)\equiv
x_i(t_1,t_2,t_3,\ldots)$ as the solution of the trigonometric
Calogero--Moser hierarchy with initial condition $(x_0,y_0)\in
\mathbb{C}^{2N}$.
\begin{theorem}\label{theorem 1} The tau function $\tau_{(X,Z-I,Z)}(x+t_1,t_2,t_3,\ldots)$
as given in \eqref{2.16}, with $(X,Z)\in C_N^{\rm trig}$ and $X$
diagonalizable, is (up to an inessential exponential factor) a
trigonometric polynomial in the variable $x$
\begin{gather}\label{3.9}
\tau_{(X,Z-I,Z)}(x+t_1,t_2,t_3,\ldots)=\prod_{i=1}^{N}2\sinh
\frac{(x-x_i(t_1,t_2,t_3,\ldots))}{2},
\end{gather}
where $x_i(t_1,t_2,t_3,\ldots)$ denotes the solution of the
trigonometric Calogero--Moser hierarchy with initial condition
$(x_0,y_0)\in\mathbb{C}^{2N}$ specified by~\eqref{3.3} as in Lemma
{\rm \ref{lemma 1}} above.
\end{theorem}
\begin{proof} Let us consider the solution $x_i(t)$ of the trigonometric Calogero--Moser hierarchy with
initial condition $(x_0,y_0)\in\mathbb{C}^{2N}$ specif\/ied
by~\eqref{3.3}, i.e.\ $U^{-1}XU=K^2(x_0)\equiv K_0^2,
U^{-1}ZU=L(x_0,y_0)\equiv L_0$ (remember from the proof of Lemma
\ref{lemma 1} that necessarily $x_i(0)\neq x_j(0)$, $\forall\,
i\neq j$).  From Proposition~\ref{proposition 2} and the
def\/inition of $\varphi_k$ in \eqref{3.8}, it follows easily that
the quantities $e^{x_i(t)}$ are the eigenvalues of the matrix
\begin{gather*}
K_0^2\exp\Biggl\{\sum_{k=1}^{\infty}t_k
\nabla\varphi_k(L_0)\Biggr\}
=K_0^2\exp\Biggl\{\sum_{k=1}^{\infty}t_k\big((L_0-I)^k-L_0^k\big)\Biggr\}.
\end{gather*}
From this, we deduce
\begin{gather*}
\prod_{i=1}^{N}e^{\frac{x+x_i(t)}{2}}\;2\sinh
\frac{(x-x_i(t))}{2}= \prod_{i=1}^N(e^{x}-e^{x_i(t)})
\\ \qquad
=\det
\Biggl\{e^{x}I-K_0^2\exp\Biggl\{\sum_{k=1}^{\infty}t_k((L_0-I)^k-L_0^k)\Biggr\}\Biggr\}
\\ \qquad
=e^{Nx}\det
\Biggl\{I-K_0^2\exp\Biggl\{-(t_1+x)I+\sum_{k=2}^{\infty}t_k((L_0-I)^k-L_0^k)\Biggr\}\Biggr\}.
\end{gather*}
Since the determinant of a matrix is invariant under conjugation,
we obtain
\begin{gather*}
\prod_{i=1}^{N}e^{\frac{x_i(t)\!-\!x}{2}}\prod_{i=1}^{N}2\sinh
\frac{(x\!-\!x_i(t))}{2} \!=\!\det
\Biggl\{I\!-\!X\exp\Biggl\{-(t_1\!+\!x)I\!+\!\sum_{k=2}^{\infty}\!t_k((Z-I)^k\!-\!Z^k)\Biggr\}\Biggr\}.
\end{gather*}
With account of \eqref{2.16}, this establishes \eqref{3.9}, up to
the inessential (in the sense that it leads to the same solution
of the KP hierarchy) exponential factor
$e^{\sum\limits_{i=1}^N\frac{x_i(t)-x}{2}}$. The proof is
complete.
\end{proof}
%%%%%%%%%%%%%%%%%%%%%%%%%%%%%%%%%%%%%%%%%%%%%%%%%%%%%%%%%%%%%%%%%%%%%%%%%%%%%%%%%%%%%%%%%%%%%%%%%%%%%%%%%%%%% Section 4
\section{The bispectral property of the KP trigonometric solitons}
The adelic Grassmannian ${\rm Gr}^{\rm ad}$ consists of the spaces
$W\in {\rm Gr}^{\rm rat}$ for which the curve $\mathop{\rm
Spec}(A_W)$ (with $A_W$ as in \eqref{2.2}) is unicursal, that is
the birational isomorphism $\mathbb{C}\to\mathop{\rm Spec}(A_W)$
corresponding to the inclusion $A_W\subset\mathbb{C}[z]$ is
bijective. As established in \cite{W1}, ${\rm Gr}^{\rm ad}$
parametrizes all commutative rank~1 algebras $\mathcal{A}$ of
bispectral dif\/ferential operators, in the sense of~\eqref{1.2}
and~\eqref{1.3}. The corresponding algebras $\mathcal{A}$ are
isomorphic to $A_W$. The joint eigenfunction of the operators in
$\mathcal{A}$ is given by the stationary Baker--Akhiezer function
$\psi_W(x,z)$ of $W$, and it can be written in terms of a pair of
matrices $(X,Z)$ satisfying $\mbox{rank}([X,Z]+I)=1$ as in
\eqref{1.4}. Since in Section~2 we have chosen in the def\/inition
of ${\rm Gr}$ the circle $S^1$ to be of radius $1$, all the
eigenva\-lues of $Z$ (which correspond to the singular points of
$\mathop{\rm Spec}(A_W)$ under the map $\mathbb{C}\to\mathop{\rm
Spec}(A_W$)) will be inside of $S^1$, see \cite{W2}. We shall thus
assume without loss of generality that any pair $(X,Z)\in C_N$ as
def\/ined in \eqref{1.1} satisf\/ies the condition that the
spectrum of $Z$ is inside the unit circle\footnote[3]{In
\cite{W1,W2}, the radius of $S^1$ is allowed to vary in the
def\/inition of ${\rm Gr}^{\rm ad}$. As explained in \cite{SW},
there is no loss of generality in f\/ixing the radius to be $1$,
since the scaling transformations $\psi_{R_\lambda
W}(x,z)=\psi_W(\lambda x,\lambda^{-1}z),\;0<\vert\lambda\vert\leq
1$, act on ${\rm Gr}$ as def\/ined in Section 2.}.

Following \cite{HI}, starting from any $W\in {\rm Gr}^{\rm ad}$
and its corresponding tau function $\tau_W(t_1, t_2,\ldots)$, we
build a function $\psi(n,t,z)=\psi(n,t_1,t_2,\ldots,z)$ via the
formula
\begin{gather}\label{4.1}
\psi(n,t,z) \!=\!(1\!+\!z)^n\exp(t,z)
\frac{\tau_W(t_1\!+\!n\!-\!{1}/{z},t_2\!-\!{n}/{2}\!-\!{1}/({2z^2}),t_3\!+\!{n}/{3}\!-\!{1}/({3z^3}),\ldots)}
{\tau_W(t_1+n,t_2-{n}/{2},t_3+{n}/{3},\ldots)}.
\end{gather}
We def\/ine a corresponding f\/lag of subspaces in
$L^2(S^1,\mathbb{C})$
\begin{gather*}
\mathcal{V}\colon \cdots\subset V_{n+1}\subset V_{n}\subset
V_{n-1}\subset \cdots
\end{gather*}
with $V_n$ the closure in $L^2(S^1,\mathbb{C})$ of the space
\begin{gather*}
V_n^{\rm alg}=\mbox{span
of}\;\{\psi(n,0,z),\psi(n+1,0,z),\psi(n+2,0,z),\ldots\}.
\end{gather*}
The set of these f\/lags was called the adelic f\/lag manifold in
\cite{HI}, and we shall denote it by ${\rm Fl}^{\rm ad}$. In order
to formulate the main result of \cite{HI}, we need to introduce
the following algebra
\begin{gather}
A_{\mathcal{V}}=\{\mbox{rational functions}\;f(z)\;\mbox{with
poles only at}\;z=-1\;\mbox{and}\;z=\infty,\nonumber
\\
\phantom{A_{\mathcal{V}}={}}{} \mbox{such that}\;\exists\;k
\in\mathbb{Z},\;\mbox{for which}\;f(z).\,V_{n}\subset
V_{n+k},\;\forall\;n\}.\label{4.2}
\end{gather}
It can be shown that the curve $\mathop{\rm
Spec}(A_{\mathcal{V}})$ is also unicursal, and that there is a
bijective birational isomorphism $\mathbb{C}\setminus\{-1\}\to
\mathop{\rm Spec}(A_{\mathcal{V}})$ which sends $-1$ and $\infty$
to two smooth points completing the curve (see \cite[Theorem
4.4]{HI}).

\begin{theorem} \label{theorem 2} \emph{(see Haine--Iliev \cite{HI})}.
Any $\mathcal{V}\in {\rm Fl}^{\rm ad}$ gives rise to a rank one
bispectral commutative algebra of difference operators
$\mathcal{A}$ as in \eqref{1.6} and \eqref{1.7}, isomorphic to
$A_{\mathcal{V}}$ as defined in \eqref{4.2}.
\end{theorem}
The function $\psi_\mathcal{V}(n,z)\equiv\psi(n,0,z)$ is the joint
eigenfuntion of the operators in $\mathcal{A}$, and is called the
(stationary) Baker--Akhiezer function of the f\/lag. From the
def\/inition of ${\rm Fl}^{\rm ad}$ and the result of \cite{W2}
which establishes a bijection between the union of the
Calogero--Moser spaces $C_{N}, N\geq 0$, introduced in \eqref{1.1}
and ${\rm Gr}^{\rm ad}$, it is easy to deduce the following lemma.

\begin{lemma}\label{lemma 2} There is a bijection $\beta\colon  \cup_{N\geq 0}C_N\to {\rm Fl}^{\rm ad}$ given by the map
\begin{gather}\label{4.3}
(X,Z)\to \psi_\mathcal{V}(n,z)=(1+z)^n\det
\{I+(X-n(I+Z)^{-1})^{-1}(zI-Z)^{-1}\}.
\end{gather}
\end{lemma}
\begin{proof} From Sato's formula \eqref{2.1} and formula \eqref{1.4} for the (stationary)
Baker--Akhiezer function of a space $W\in {\rm Gr}^{\rm ad}$, one
deduces easily that the corresponding tau function is given~by
\begin{gather*}
\tau_W(t_1,t_2,t_3,\ldots)=\det
\Biggl\{X-\sum_{k=1}^{\infty}kt_kZ^{k-1}\Biggr\}.
\end{gather*}
Using the fact that the spectrum of $Z$ is inside the unit circle,
one f\/inds
\begin{gather}
\tau_{\mathcal{V}}(n,t_1,t_2,\ldots)\!\equiv\!
\tau_W\biggl(t_1\!+\!n,t_2\!-\!\frac{n}{2},t_3\!+\!\frac{n}{3},\ldots\biggr)
\!=\!\det
\Biggl\{X\!-\!\sum_{k=1}^{\infty}kt_kZ^{k\!-\!1}\!-\!n(I\!+\!Z)^{\!-\!1}\Biggr\}\label{4.4}.
\end{gather}
From \eqref{4.1}, a simple computation shows then that the
(stationary) Baker--Akhiezer function $\psi_{\mathcal{V}}(n,z)$ of
the corresponding adelic f\/lag is given by \eqref{4.3}, which
establishes the lemma.
\end{proof}
We can now formulate the main result of our paper.
\begin{theorem}\label{theorem 3} The stationary Baker--Akhiezer function $\psi_W(x,z)$
of a space $W\in {\rm Gr}^{\rm trig}$ as defined in \eqref{2.17},
besides being an eigenfunction of a commutative algebra of
differential operators in $x$
\begin{gather}\label{4.5}
A_f\psi_W(x,z)\equiv \sum_{\emph{finitely many}\;j\in \mathbb{N}}
a_j(x) \frac{\partial^j}{\partial
x^j}\;\psi_W(x,z)=f(z)\psi_W(x,z)\qquad\forall\;f\in A_W,
\end{gather}
is also an eigenfunction of a commutative algebra of difference
operators in the spectral variable~$z$ built from an adelic flag
$\mathcal{V}$, i.e.
\begin{gather}\label{4.6}
B_g\psi_W(x,z)\equiv\sum_{\emph{finitely
many}\;j\in\mathbb{Z}}b_j(z)\psi_W(x,z+j)
=g(x)\psi_W(x,z)\qquad\forall\;g\in A_{\mathcal{V}}.
\end{gather}
\end{theorem}
\begin{proof} Any stationary Baker--Akhiezer function associated with a space $W$
in the Segal--Wilson Grassmannian, for which $\mathop{\rm
Spec}(A_W)$ is an irreducible af\/f\/ine algebraic curve (which
completes by adding one non-singular point at inf\/inity),
satisf\/ies \eqref{4.5} for a commutative algebra of
dif\/ferential operators $A_f$, $f\in A_W$. This is a
reformulation of the classical Burchnall--Chaundy--Krichever
theory as explained in Section 6 of \cite{SW}. In particular, it
applies to any $W\in {\rm Gr}^{\rm trig}\subset {\rm Gr}^{\rm
rat}$.

From the def\/inition of ${\rm Gr}^{\rm trig}$ \eqref{2.17}, the
same proof as in Proposition~\ref{proposition 1} shows that the
stationary Baker--Akhiezer function $\psi_W(x,z)$ of $W\in {\rm
Gr}^{\rm trig}$ is given by \eqref{2.13}. Let us def\/ine
\begin{gather}
\psi^b(n,z)=\psi_W(\log(1+z),n)\label{4.7}
\\
\phantom{\psi^b(n,z)}{} =(1+z)^n\det
\big\{I-X((1+z)I-X)^{-1}(nI-Z)^{-1}\big\}.\label{4.8}
\end{gather}
We claim that $\psi^b(n,z)$ is the stationary Baker--Akhiezer
function of an adelic f\/lag. In view of the characterization
given in \eqref{4.3} of the stationary Baker--Akhiezer function of
an adelic f\/lag, using the fact that the determinants of a matrix
and its transpose are equal, it is enough to f\/ind two matrices
$\tilde{X}$ and $\tilde{Z}$ satisfying
$\mbox{rank}([\tilde{Z},\tilde{X}]+I)=1$, so that $\psi^b(n,z)$
can be put into the form
\begin{gather}\label{4.9}
\psi^b(n,z)=(1+z)^n\det
\big\{I+(zI-\tilde{Z})^{-1}(\tilde{X}-n(I+\tilde{Z})^{-1})^{-1}\big\}.
\end{gather}
It is easy to see that \eqref{4.8} agrees with \eqref{4.9} by
picking
\begin{gather}\label{4.10}
\tilde{X}=ZX^{-1}\qquad\mbox{and}\qquad \tilde{Z}=X-I.
\end{gather}
Since{\samepage
\begin{gather}\label{4.11}
[\tilde{Z},\tilde{X}]+I=[X,ZX^{-1}]+I=XZX^{-1}-Z+I,
\end{gather}
by the def\/inition \eqref{1.9} of $C_N^{\rm trig}$, we have that
$\mbox{rank}([\tilde{Z},\tilde{X}]+I)=1$, which establishes our
claim.}

Since $\psi^b(n,z)$ is the stationary Baker--Akhiezer function of
an adelic f\/lag $\mathcal{V}$, by Theorem~\ref{theorem 2}, there
exists a commutative algebra of dif\/ference operators $\{B_g,
g\in A_\mathcal{V}\}$, so that
\begin{gather*}
B_g\psi^b(n,z)\equiv\sum_{{\rm finitely \
many}\;j\in\mathbb{Z}}b_j(n)\psi^b(n+j,z)=g(z)\psi^b(n,z).
\end{gather*}
Remembering the def\/inition \eqref{4.7} of $\psi^b(n,z)$ in terms
of $\psi_W(x,z)$, we obtain
\begin{gather*}
\sum_{{\rm finitely\
many}\;j\in\mathbb{Z}}b_j(z)\psi_W(x,z+j)=g(e^{x}-1)\psi_W(x,z)\qquad
\forall\;g\in A_{\mathcal{V}},
\end{gather*}
which establishes \eqref{4.6} and completes the proof of the
theorem.
\end{proof}
\begin{corollary} The diagram \eqref{1.8} with $b$ and $b^{C}$ defined as in \eqref{1.11} and \eqref{1.12},
$\beta^{\rm trig}$ and $\beta$ defined respectively as in
\eqref{1.13} and \eqref{4.3}, is commutative.
\end{corollary}
\begin{proof}
Looking at the proof of Theorem~\ref{theorem 3} and more
specif\/ically at equations \eqref{4.7}, \eqref{4.10}
and~\eqref{4.11}, we have in fact constructed a map, the inverse
of the map $b^{C}$ in \eqref{1.12}
\begin{gather} \label{4.12}
(b^C)^{-1}(X,Z)=\big((X^{-1})^tZ^t, X^t-I\big),
\end{gather}
with the property that
\begin{gather*}
\psi_{\beta^{\rm
trig}(X,Z)}(x,z)=\psi_{\beta((b^C)^{-1}(X,Z))}(z,e^x-1)
=\psi_{b(\beta((b^C)^{-1}(X,Z)))}(x,z).
\end{gather*}
This establishes the commutativity of the diagram \eqref{1.8}.
\end{proof}
\begin{remark}
In \cite{K} (see also \cite{R} for KdV solitons), it was shown
that for any $W\in {\rm Gr}^{\rm rat}$, the Baker--Akhiezer
function $\psi_W(x,z)$, besides satisfying \eqref{4.5}, also
satisf\/ies
\begin{gather*}
\sum_{j=1}^{M}b_j(z,\partial/\partial
z)\psi_W(x,z+\lambda_j)=g(x)\psi_W(x,z),
\end{gather*}
for appropriate $\lambda_j\in\mathbb{C}$, with
$b_j(z,\partial/\partial z)$ ordinary dif\/ferential operators in
$z$. As shown in Theorem~\ref{theorem 3}, the KP trigonometric
solitons are distinguished by the property that this equation can
be taken to be a dif\/ference equation in $z$, with the $b_j$'s
functions of $z$ and the $\lambda_j$'s integers.
\end{remark}
%%%%%%%%%%%%%%%%%%%%%%%%%%%%%%%%%%%%%%%%%%%%%%%%%%%%%%%%%%%%%%%%%%%%%%%%%%%%%%%%%%%%%%%%%%%%%%%%%%%%%%%%%%%%% Section 5
\section{Ruijsenaars' duality revisited}
A simple computation, using the commutativity of the diagram
\eqref{1.8}, the def\/inition of $(b^C)^{-1}$ in \eqref{4.12} and
formula \eqref{4.4} for the tau function of an adelic f\/lag,
shows that for $(X,Z)\in C_N^{\rm trig}$
\begin{gather} \label{5.1}
\tau_{b^{-1}(\beta^{\rm trig}(X,Z))}(n,t_1,t_2,\ldots)=(-1)^N(\det
X)^{-1}\det
\Biggl\{nI-Z+\sum_{k=1}^{\infty}kt_k(X-I)^{k-1}X\Biggr\}.
\end{gather}
To make contact with the rational Ruijsenaars--Schneider system,
we assume that $Z$ is diagonalizable, with no two eigenvalues
dif\/fering by a unit. Then, by an argument similar to the one
used in Lemma~\ref{lemma 1}, there is a conjugation
\begin{gather} \label{5.2}
T^{-1}ZT=K^{\rm RS}(\lambda),\qquad T^{-1}XT=L^{\rm
RS}(\lambda,\theta),
\end{gather}
which diagonalizes $Z$ in such a way that
\begin{gather} \label{5.3}
K^{\rm RS}_{ij}=\lambda_i\delta_{ij}, \quad
\mbox{with}\;\lambda_i\neq \lambda_j,\;\forall \,i\neq j,\quad
\mbox{and} \quad L^{\rm
RS}_{ij}=\frac{\mu_i\mu_j}{1+\lambda_j-\lambda_i},\;\quad\mbox{with}\;\mu_i\neq
0.
\end{gather}
Putting
\begin{gather} \label{5.4}
\mu_i=e^{{\theta_i}/{2}}\prod_{k\neq
i}\biggl[1-\frac{1}{(\lambda_i-\lambda_k)^2}\biggr]^{1/4},
\end{gather}
the rational Ruijsenaars--Schneider system \cite{RS} is the
Hamiltonian system
\begin{gather}\label{5.5}
\dot{\lambda}_i=\frac{\partial H}{\partial \theta_i},\qquad
\dot{\theta_i}=-\frac{\partial H}{\partial \lambda_i},
\end{gather}
with Hamiltonian $H(\lambda,\theta)=\mathop{\rm tr}L^{\rm
RS}(\lambda,\theta)$. It is a completely integrable system, whose
solution (as well as the solution of the f\/lows commuting with
it) is described by the next proposition.
\begin{proposition} \label{proposition 3} \emph{(\cite{RS},
see also \cite[Theorem 27.5]{Su} for a proof)}. Let $\varphi\colon
gl(N,\mathbb{C})\to\mathbb{C}$ be an {\rm Ad}-invariant function.
Then, the solution of the system \eqref{5.5} with Hamiltonian
$H(\lambda,\theta)=\varphi\big(L^{\rm RS}(\lambda,\theta)\big)$
and initial conditions
$(\lambda_0,\theta_0)=\big(\lambda_1(0),\ldots,
\lambda_N(0),\theta_1(0),\ldots,\theta_N(0)\big)\in\mathbb{C}^{2N}$
satisfying $\lambda_i(0)-\lambda_j(0)\notin\{0,1\},\;\forall \,
i\neq j$, is such that the functions $\lambda_i(t)$ are the
eigenvalues of the matrix
\begin{gather*}
K^{\rm RS}_0+tL^{\rm RS}_0\nabla \varphi\big(L^{\rm RS}_0\big),
\end{gather*}
with $K^{\rm RS}_0=K^{\rm RS}(\lambda_0)$, $L^{\rm RS}_0=L^{\rm
RS}(\lambda_0,\theta_0)$ as in \eqref{5.3} and \eqref{5.4}, and
$\nabla\varphi$ the gradient of $\varphi$.
\end{proposition}
Let us now consider the following commuting Hamiltonians
$H_k(\lambda,\theta)=\varphi_k(L^{\rm RS}(\lambda,\theta))$, with
\begin{gather}\label{5.6}
\varphi_k(L^{RS}(\lambda,\theta))=-\mathop{\rm tr}\{L^{\rm
RS}(\lambda,\theta)-I\}^k,\qquad k=1,2,\ldots,
\end{gather}
and def\/ine accordingly $\lambda_i(t_1,t_2,\ldots)$ to be the
solution obtained by f\/lowing along these various Hamiltonian
vector f\/ields, starting from some initial condition
$(\lambda_0,\theta_0)\in\mathbb{C}^{2N}$ as in
Proposition~\ref{proposition 3}. As before, we refer to
$\lambda_i(t_1,t_2,\ldots)$ as the solution of the rational
Ruijsenaars--Schneider hierarchy with initial condition
$(\lambda_0,\theta_0)\in\mathbb{C}^{2N}$. Our f\/inal theorem,
when  combined with Theorem~\ref{theorem 1}, reveals a duality
between the trigonometric Calogero--Moser hierarchy and the
rational Ruijsenaars--Schneider hierarchy, in terms of the
bispectral map $b$ def\/ined in \eqref{1.11}. This duality was
f\/irst discovered by Ruijsenaars in \cite{R1}, by studying the
scattering theory of these systems.
\begin{theorem} \label{theorem 4} Let $W=\beta^{\rm trig}(X,Z)\in {\rm Gr}^{\rm trig}$, with $(X,Z)\in C_N^{\rm trig}$
a trigonometric Calogero--Moser pair such that $Z$ is
diagonalizable with no two eigenvalues differing by a unit. Then
\begin{gather}\label{5.7}
\tau_{b^{-1}(W)}(n,t_1,t_2,\ldots)= (-1)^N
e^{-\sum\limits_{i=1}^{N}\theta_i(0)}\prod_{i=1}^N(n-\lambda_i(t_1,t_2,\ldots)),
\end{gather}
where $\lambda_i(t_1,t_2,\ldots)$ is the solution of the rational
Ruijsenaars--Schneider hierarchy, with initial condition
$(\lambda_0,\theta_0)$ defined by \eqref{5.2}, \eqref{5.3} and
\eqref{5.4}.
\end{theorem}
\begin{proof} Let $L^{\rm RS}(\lambda_0,\theta_0)\equiv L^{\rm RS}_0$ and $K^{\rm RS}(\lambda_0)\equiv K^{\rm RS}_0$
be def\/ined from $X$ and $Z$ as in \eqref{5.2}, \eqref{5.3} and
\eqref{5.4}. Since a determinant is invariant under conjugation,
using the def\/inition of $\varphi_{k}$ \eqref{5.6}, from
\eqref{5.1} we f\/ind
\begin{gather*}
\tau_{b^{-1}(W)}(n,t_1,t_2,\ldots) =(-1)^N\big(\det L^{\rm
RS}_0\big)^{-1} \det \Biggl\{nI-K^{\rm RS}_0-
\sum_{k=1}^{\infty}t_k \;L^{\rm RS}_0\nabla\varphi_k\big(L^{\rm
RS}_0\big)\Biggr\}.
\end{gather*}
By Cauchy's determinant formula \eqref{2.14}, using \eqref{5.4},
we get
\begin{gather*}
\big(\det L^{\rm RS}_0\big)^{-1}=\prod_{i=1}^{N}e^{-\theta_i(0)}.
\end{gather*}
From Proposition~\ref{proposition 3}, it follows that the
functions $\lambda_i(t_1,t_2,\ldots)$ which solve the rational
Ruijse\-naars--Schneider hierarchy with initial condition
$(\lambda_0,\theta_0)$ are the eigenvalues of the matrix
\begin{gather*}
K^{\rm RS}_{0}+\sum_{k=1}^{\infty}t_k L^{\rm
RS}_0\nabla\varphi_k\big(L^{\rm RS}_0\big),
\end{gather*}
which establishes the assertion \eqref{5.7} and completes the
proof of the theorem.
\end{proof}
\begin{remark} A trigonometric Calogero--Moser pair $(X,Z)$ for which both $X$ and $Z$
are diagonalizable with no two eigenvalues of $Z$ dif\/fering by a
unit, can be represented (modulo conjugation) by
$\big(K^2(x_0),L(x_0,y_0)\big)$ as in \eqref{3.3}, or $\big(L^{\rm
RS}(\lambda_0,\theta_0),K^{\rm RS}(\lambda_0)\big)$ as in
\eqref{5.2}. Let $W=\beta^{\rm trig}(X,Z)$ be the corresponding
space in ${\rm Gr}^{\rm trig}$. By Theorem~\ref{theorem 1}, the
zeros of $\tau_W(x+t_1,t_2,\ldots)$ as a function of $x$ solve the
trigonometric Calogero--Moser hierarchy with initial condition
$(x_0,y_0)$ and, by Theorem~\ref{theorem 4}, the zeros as a
function of $n$ of $\tau_{b^{-1}(W)}(n,t_1,t_2,\ldots)$ solve the
rational Ruijsenaars--Schneider hierarchy with initial condition
$(\lambda_0,\theta_0)$. To describe a repulsive interaction in the
trigonometric Calogero--Moser system \eqref{3.1}, one chooses
the~$x_i$'s real with $x_1<\cdots<x_N$, the~$y_i$'s imaginary and
time imaginary too. In this case $L(x,y)$ is anti-hermitian and
thus diagonalizable with imaginary eigenvalues. As shown in
\cite[Section 2C]{R1}, the matrix $T$ in \eqref{5.2} which
conjugates the pair $\big(K^2(x),L(x,y)\big)$ to $\big(L^{\rm
RS}(\lambda,\theta),K^{\rm RS}(\lambda)\big)$, is then uniquely
determined by requiring it to be unitary and such that
$\sqrt{-1}\,\lambda_1<\cdots<\sqrt{-1}\,\lambda_N$ and $\mu_i>0$.
The real solution of the hyperbolic Calogero--Moser system is then
given by
$\big(x_i(\sqrt{-1}\,t),\sqrt{-1}\,y_i(\sqrt{-1}\;t)\big),t\in\mathbb{R}$,
and the map
$\big(\sqrt{-1}\;y_i,x_i\big)\to\big(\sqrt{-1}\,\lambda_i,\theta_i\big)$,
is the scattering map, providing the action-angle variables for
the system. Vice versa, the map
$\big(\theta_i,\sqrt{-1}\,\lambda_i\big)\to\big(x_i,\sqrt{-1}\,y_i\big)$
gives the action-angle variables for the rational
Ruijsenaars--Schneider system.
\end{remark}
%%%%%%%%%%%%%%%%%%%%%%%%%%%%%%%%%%%%%%%%%%%%%%%%%%%%%%%%%%%%%%%%%%%%%%%%%%%%%%%%%%%%%%%%%%%%%%%%%%%%%% Acknowledgements

\subsection*{Acknowledgements}
I wish to thank S.N.M. Ruijsenaars for his comments about
\cite{HI} during the `International Workshop on Special Functions,
Orthogonal Polynomials, Quantum Groups and Related Topics'
dedicated to Dick Askey 70th birthday (Bexbach, October 2003),
which hinted at some of the results presented here, as well as for
sending \cite{R2}. I also thank two anonymous referees for
stimu\-lating suggestions, which led to improvement of the f\/inal
form of the paper.  Partial support from the European Science
Foundation Programme MISGAM, the Marie Curie RTN ENIGMA and a
Grant of the Belgian National Science Foundation (FNRS) are also
gratefully acknowledged.

\pdfbookmark[1]{References}{ref}
\LastPageEnding

\end{document}